\documentclass[aps,twocolumn,pre,nopacs,balancelastpage,amssymb,groupedaddress]{revtex4}
\usepackage{epsfig}
\usepackage{verbatim}	 
\usepackage{amssymb}
\usepackage{bbold}
\usepackage{amsmath}
\usepackage{psfrag}
\usepackage{color}
\usepackage{ulem}
\newcommand{\bi}{\bibitem}

\usepackage[latin1]{inputenc}
\usepackage[english]{babel}
\usepackage{graphicx}
\usepackage{amscd}
\usepackage{eucal}
\usepackage{bm}
\usepackage{float}
\input{epsf}
\begin{document}

\title{Eigenstate thermalization and ensemble equivalence in few-body fermionic systems}
\author{Ph.~Jacquod}

\affiliation{Department of Quantum Matter Physics, University of Geneva, CH-1211 Geneva, Switzerland \\ School of Engineering, University of Applied Sciences of Western Switzerland HES-SO CH-1951 Sion, Switzerland }

\date{\today}

\begin{abstract}
  We investigate eigenstate thermalization from the point of view of vanishing
  particle and heat currents between a few-body fermionic
  Hamiltonian prepared in one of its eigenstates and an external, weakly coupled Fermi-Dirac gas. The 
  latter acts as
  a thermometric probe, with its temperature and chemical potential 
  set so that there is neither particle nor
  heat current between the two subsystems. We argue that 
  the probe temperature can be attributed to the 
  few-fermion eigenstate in the sense that (i) it varies smoothly with energy from eigenstate to eigenstate,  
  (ii) it is equal to the temperature obtained from
a thermodynamic relation in a wide energy range, 
  (iii) it is independent of details of the 
  coupling between the two systems in a finite parameter range, (iv) it satisfies the transitivity 
condition underlying the zeroth law of 
thermodynamics and (v) it is consistent with Carnot's theorem.
These conditions are essentially independent of the interaction between the few fermions.
When the interaction strength is weak, however, orbital occupancies in the few fermion system
differ from the Fermi-Dirac distribution so that partial currents from or to the 
probe will eventually change its state.
We  find that (vi) above a certain critical interaction strength, 
orbital occupancies become close to the Fermi-Dirac distribution, leading to a true 
equilibrium between the few-fermion system and the probe. 
We conclude that for few-body systems with sufficiently strong interaction,
the eigenstate thermalization hypothesis of Deutsch and Srednicki
is complemented by ensemble equivalence: individual many-body eigenstates
define a microcanonical ensemble that is equivalent to a canonical ensemble. 
\end{abstract}

\maketitle

\section{Introduction}

In equilibrium statistical mechanics, observable properties of macroscopic systems are given
by ensemble averages over microscopic states. 
When considering the microcanonical ensemble, the average is taken over states
of similar energy. All states in that average contribute with equal weight -- this 
is the postulate of equal a priori probabilities~\cite{Huang}. The process following which 
almost all initial out-of-equilibrium conditions evolve into equilibrium states well represented by the microcanonical ensemble
is called thermalization. Within classical mechanics, thermalization and the emergence of 
equilibrium statistical mechanics is standardly
explained at a microscopic level by dynamical complexity~\cite{Lili} : regardless of their origin, 
almost all classical trajectories of chaotic dynamical systems eventually look 
the same as they explore ergodically the constant energy hypersurface in phase space. It is usually accepted that
the macroscopic laws of classical statistical mechanics emerge from this ergodicity~\cite{Leb93}. 

The situation is different in quantum mechanics, whose time-evolution cannot by itself lead to the uniform
covering of the constant energy manifold typical of the microcanonical measure. Instead it has been postulated
that thermalization occurs at the level of individual eigenstates~\cite{Deu91,Sre94}, in such a way that
expectation values of almost all observables
taken over almost any such state in a narrow energy interval give the same result in the thermodynamic limit, tending moreover 
to the microcanonical average~\cite{Deu91,Sre94,Rig08}. For recent reviews of and details on this
{\it Eigenstate Thermalization Hypothesis} (ETH), see Refs.~\cite{Deu18,Ale16}. 

A question that naturally arises is whether the ETH is accompanied by 
an ensemble equivalence similar to the one prevailing in 
statistical mechanics~\cite{Huang}. In the spirit of the standard construction of the
canonical ensemble from the microcanonical one,  
it has been shown that 
tracing over some of the degrees of freedom of a pure quantum state gives, for almost all pure states in a narrow energy
interval, a reduced density matrix corresponding
to the canonical ensemble~\cite{Tas97,Gol06,Pop06}. 
A similar procedure showed that typical many-body eigenstates have consistently defined
thermodynamic entropies~\cite{Deu10}.
Taken together with the ETH, this can be interpreted as ensemble 
equivalence at the level of individual many-body eigenstates. This equivalence 
requires however a bipartitioning of quantal eigenstates and one may wonder if a temperature 
can be attributed to individual eigenstates without partitioning.
To the best of our knowledge, 
this question was first asked in Ref.~\cite{Fla97}, where an ad hoc microcanonical partition function was 
constructed from the shape of many-body eigenstates and shown to bear similarities with the
canonical partition function. In particular, the resulting occupation number $f(E)$ of single-particle
orbitals was found to become a smooth,
monotonously decreasing function of the energy $E$ for sufficiently strong perticle-particle 
interactions~\cite{Fla97,Bor16}.

\begin{figure}
\centering
\includegraphics[width=0.85 \columnwidth]{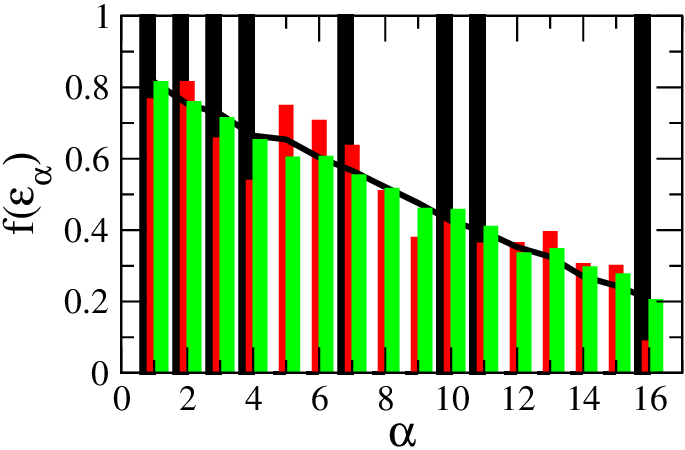}
\caption{\label{fig:fig1} (Color online) Occupation number for the 1001$^{\rm st}$ many-body eigenstate
of ${\cal H}_{\rm sys}$ \eqref{eq:tbrim} with $m=16$ orbitals and $n=8$ particles, for $U/\Delta=0.$ [black histogram,
with temperature and chemical potential $(T/\Delta,\mu/\Delta)=(5.41,0.0141)$]
0.1 [red, $(T/\Delta,\mu/\Delta)=(5.43,0.0139)$] and 0.2 [green, 
$(T/\Delta,\mu/\Delta)=(5.47,0.0135)$]
The black solid line is the Fermi function corresponding to $(T/\Delta,\mu/\Delta)=(5.47,0.0135)$. 
It is slightly unsmooth because the one-body spectrum of \eqref{eq:tbrim} is not equidistant.}
\vspace{-0.4cm}
\end{figure}

Motivated by this issue and inspired by the experimental technique of scanning thermal 
microscopy~\cite{Maj99}, 
we show in this manuscript that a temperature consistant with 
a number of standard thermodynamic criteria can indeed 
be attributed to individual
many-body eigenstates $|A\rangle$ of few-fermion systems, by coupling them weakly
to an external Fermi-Dirac gas. 
The coupling allows for particle exchange between the two systems,
and the external Fermi-Dirac gas acts as a  
probe~\cite{But86,Dub09,Cas10,Mea14,Jac10,Sta16}:
its temperature $T_A$ and chemical potential $\mu_A$ are set so that particle and heat currents between 
the few-fermion system, prepared in its eigenstate $|A\rangle$, and the probe vanish. We find
that $T_A$ varies smoothly with energy from eigenstate to eigenstate regardless of the strength
of the particle-particle interaction, despite the fact that the latter strongly influences the structure of the 
many-body eigenstates (see Fig.~\ref{fig:fig1}).
In a large energy range, the obtained temperature is furthermore
well approximated by the thermodynamic relation~\cite{Huang} 
\begin{equation}\label{eq:temperatureS}
T=(\partial S/\partial E)^{-1}_X \, ,
\end{equation}
with the entropy $S$ determined by the density of states $\rho(E)$ through Boltzmann's formula
\begin{equation}\label{eq:entropy}
S(E) = k_{\rm B} \ln \rho(E) \, ,
\end{equation}
and the subscript $X$ indicating the set of thermodynamic variables kept constant. In our case, 
this is the system's volume, via the fixed number of orbitals considered. 

It is tempting to attribute the probe temperature $T_A$ to the
many-body eigenstate of the few-fermion system to which the probe is connected. 
Below, we argue that this thermometric definition of eigenstate temperature 
is indeed consistent with standard thermodynamic definitions in the sense that 
(i) it varies smoothly from eigenstate to eigenstate and is monotonously increasing with energy,
(ii) it is equal to a temperature independently obtained from the thermodynamic relation of 
Eqs.~\eqref{eq:temperatureS} and \eqref{eq:entropy} in a large energy range, 
(iii) it is independent of details of the system-probe coupling in a finite parameter
range, (iv) it satisfies the transitivity 
condition underlying the zeroth law of 
thermodynamics and (v) it is consistent with Carnot's theorem.
Quite interestingly, these consistency 
conditions are valid regardless of the particle-particle interaction in the few fermion system.
We argue that ensemble equivalence is however achieved only once a further condition is imposed, 
that (vi) all partial currents between single-particle orbitals in the few fermion system and the Fermi-Dirac gas
vanish. This latter condition ensures that the few fermion state does not eventually change over time, and
guarantees that a true equilibrium exists between the two subsystems. 
We find that condition (vi) is achieved for sufficiently strong particle-particle interaction,
$U \gtrsim U_c \sim n^{-3}$, with the number $n$ of fermions, in systems with fixed particle density.

Our numerical results are based
on exact diagonalization of few-particle
systems, and are therefore limited to systems with up to $n=8$ fermions. They suggest that the
 threshold interaction strength $U_c$ above which (iv) is valid, is parametrically similar to the 
 many-body quantum chaos threshold derived in Refs.~\cite{Abe90,Jac97}, giving a 
 parametric critical interaction strength going down algebraically with the system size and 
 the number of fermions. 
Taken together, the conditions (i)--(vi) suggest that 
small fermionic systems with sufficient but not too strong
interaction have many-body eigenstates that not only satisfy the ETH -- this was already known -- 
but that each of them defines both a microcanonical and a canonical ensemble which are equivalent to each other
in the standard thermodynamic sense. 

\section{The model}

We consider systems of $n$ interacting, spinless fermions with Hamiltonian given by
the two-body random ensemble~\cite{Bro81}
\begin{equation}\label{eq:tbrim}
{\cal H}_{\rm sys} = \sum \epsilon_\alpha c^\dagger_\alpha c_\alpha + \sum U_{\alpha,\beta}^{\gamma,\delta} c^\dagger_\alpha c_\beta
c_\gamma c_\delta \, .
\end{equation}
Here $c^\dagger_\alpha$ and $c_\alpha$ are creation and annihilation operators obeying
fermionic anticommutation relations and $\epsilon_\alpha \in [-m \Delta/2, m \Delta /2] $ are $m$ single-particle orbital energies with average spacing $\Delta$. We take $\epsilon_\alpha$ as eigenvalues of a $m \times m$ 
random matrix of the Gaussian orthogonal ensemble~\cite{Bro81}. 
Fermions occupying these single-particle energies interact via a two-body 
interaction with matrix elements randomly distributed as $U_{\alpha,\beta}^{\gamma,\delta}  \in [-U,U]$.
The total number of many-body eigenstates
of ${\cal H}_{\rm sys}$ is $N=m!/n!(m-n)!$ and the corresponding eigenvalues are distributed
over a bandwidth $B \simeq n(m-n) \Delta$.
In this manuscript
we focus on systems at half-filling with $n=m/2$ fermions.
Many-body eigenstates of ${\cal H}_{\rm sys}$ are linear combinations of totally antisymmetric 
$n$-body wavefunctions over $m$ orbitals. Each such many-body eigenstate $|A\rangle$ has single-orbital
occupancies $f_A(\epsilon_\alpha) \in [0,1]$. Occupancies for many-body states corresponding to three different 
interaction strengths are shown in Fig.~\ref{fig:fig1}.

We weakly couple that system to an external noninteracting 
fermionic gas with Hamiltonian
\begin{equation}\label{eq:FD}
{\cal H}_{\rm FD} = \sum E_i d^\dagger_i d_i \, .
\end{equation}
In this Gedankenexperiment, we assume that $n$ consecutive energy eigenvalues $E_i$, $i=i_0, i_0+1, ... i_0+n-1$,
are equal to the $n$ single-particle energies $\epsilon_\alpha$, $\alpha=1,2,... n$. 
Furthermore, that external fermionic gas is thermalized in the standard textbook way, being e.g. connected to an 
infinite external reservoir at a tunable  temperature $T$. Consequently, we assume that
it is in a mixed state where its single-particle occupancies obey a Fermi-Dirac distribution 
\begin{equation}
f_{\rm FD}(E_i,\mu,T)=\frac{1}{1+\exp((E_i-\mu)/ k_{\rm B} T)}
\end{equation}
with chemical potential $\mu$. 
This Fermi-Dirac gas (FDG) is the probe which will  attribute a temperature to each many-body  
eigenstates of the few interacting fermion system of Eq.~\eqref{eq:tbrim}. 

We assume that the tunneling 
amplitude $t$ between 
the FDG probe and the few interacting fermion system is small, constant and strictly energy conserving.
The few-fermion system is prepared in one of its many-body eigenstates $|A\rangle$.
Under our assumptions, the probe-system coupling induces particle ($I_A$) and heat ($J_A$)
currents between the two subsystems. We approximate them as sums over energy-conserving
partial currents,
\begin{subequations}\label{eq:currents}
\begin{eqnarray}
I_A &=& t^2 \sum_\alpha [f_A(\epsilon_\alpha) - f_{\rm FD}(E_i=\epsilon_\alpha,\mu,T)]\\
J_A &=& t^2 \sum_\alpha  (\epsilon_\alpha-\mu) [f_A(\epsilon_\alpha) - f_{\rm FD}(E_i=\epsilon_\alpha,\mu,T)]\, . \qquad 
\end{eqnarray}
\end{subequations}
For each many-body eigenstate $|A\rangle$, $I_A$
and $J_A$ are functions of $T$ and $\mu$.  We then
tune the temperature $T \rightarrow T_A$ and chemical potential $\mu \rightarrow \mu_A$ of the FDG to ensure that 
$I_A(\mu_A,T_A)=J_A( \mu_A,T_A) \equiv 0$. If a temperature can at all be attributed
to $|A\rangle$, then this temperature is $T_A$~\cite{Huang}. 

\begin{figure}
\centering
\includegraphics[width=1. \columnwidth]{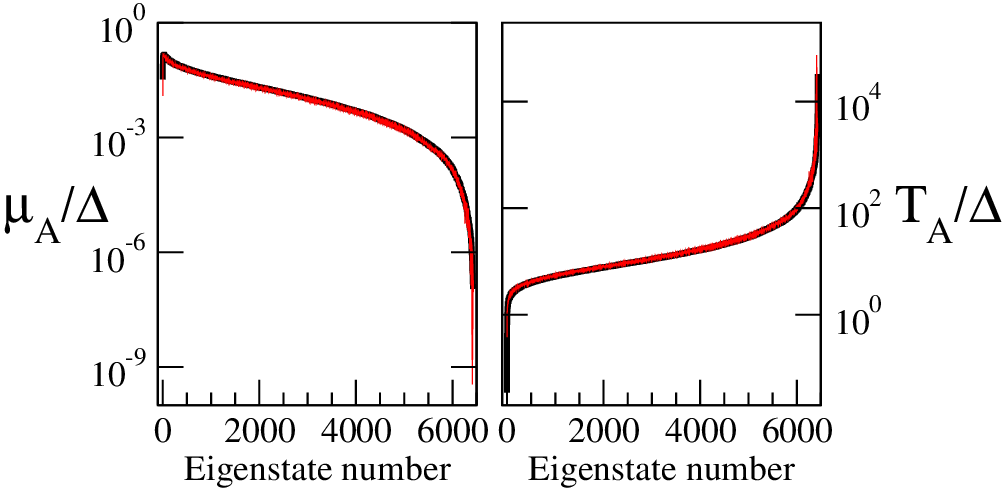}
\caption{\label{fig:fig2} (Color online) Chemical potential (left) and temperature (right) vs. many-body eigenstate 
number $A \in [1,N/2]$ for a single realization of the two-body randomly interacting fermion model \eqref{eq:tbrim} with 
$m=16$ orbitals and $n=8$ fermions and thus $N=12870$ many-body states. 
Black dots are for $U/\Delta=0.$ and red dots for $U/\Delta=0.2$. }
\vspace{-0.4cm}
\end{figure}

\section{Probe temperature and chemical potential}

Our numerical procedure is the following.
We diagonalize exactly the Hamiltonian \eqref{eq:tbrim} and calculate its full set
of many-body eigenvectors and the corresponding eigenvalues. 
We calculate the single-particle occupancies of each eigenstate and use a Newton-Raphson
algorithm to obtain the values $T_A$ and $\mu_A$ defined by $I_A=J_A=0$.

Fig.~\ref{fig:fig2} shows results for a single realization of \eqref{eq:tbrim} with $n=8$ fermions on $m=16$ orbitals, for 
$U/\Delta=0$ and  $U/\Delta=0.2$. We see that, first, both temperature and chemical potential vary smoothly from 
eigenstate to eigenstate in both cases. Second, turning on the interaction
has only a small effect on both temperature and chemical potential, despite the fact that it changes the structure
of individual many-body eigenstates very significantly, as illustrated in Fig.~\ref{fig:fig1}.

\begin{figure*}[hbt!]
\includegraphics[width=2.02 \columnwidth]{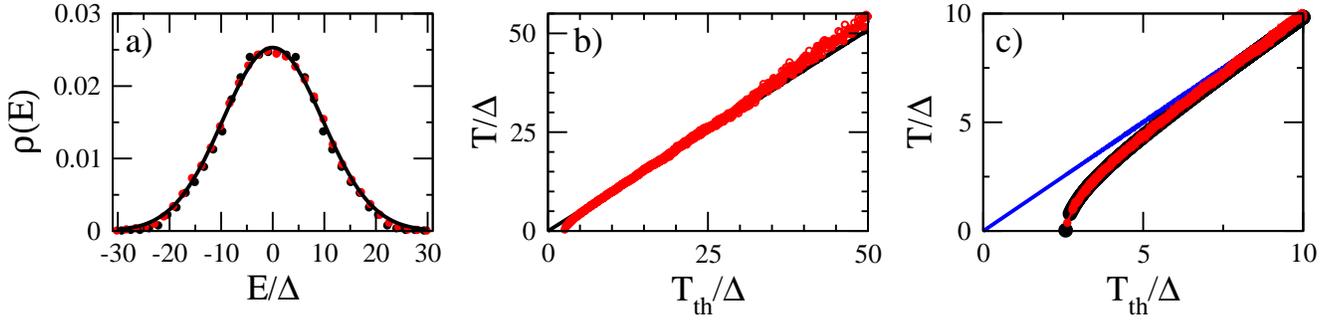}
\caption{\label{fig:fig3} (Color online) a) Many-body density of states. b) Numerically obtained 
eigenstate temperature $T$ vs. theoretical temperature $T_{\rm th}$ of Eq.~(\ref{eq:temp}) for $0<T_{\rm th}/\Delta \le 50$. The interval contains about 90\% of the positive temperature eigenstates.
c) Same as in panel b) for a restricted range $0 < T_{\rm th}/\Delta \le 10$, showing deviations at low energy due to 
non-Gaussian tails in the density of states~\cite{Bro81}. All panels correspond to a single realization of the two-body randomly interacting 
fermion model \eqref{eq:tbrim} with 
$m=16$ orbitals, $n=8$ fermions and $N=12870$ many-body states. 
Black dots are for $U=0.$ and red dots for $U=0.2 \Delta$.}
\vspace{-1cm}
\end{figure*}

We next investigate to what extent the obtained 
probe temperature corresponds to the thermodynamic relation for the few-fermion system given in
Eq.~\eqref{eq:temperatureS}.
For weak interaction, $U \ll \Delta$, the density of states
is Gaussian, $\rho(E) = \rho_0 \exp[-E^2/2 \sigma^2]$ with 
a variance $\sigma^2 \propto n (m-n)$ for dilute systems $n \ll m$~\cite{Bro81}.
With Eqs.~(\ref{eq:temperatureS}) and (\ref{eq:entropy}), this density of states gives a theoretical 
system temperature [see also Ref.~\cite{Bor16}]
\begin{equation}\label{eq:temp}
T_{\rm th} (E) = -\sigma^2/k_{\rm B} E\, .
\end{equation}
This temperature diverges in the middle of the spectrum and becomes negative at higher temperature,
as standardly happens in systems with nonmonotonous density of states~\cite{Ram56}. We focus on the 
lower half of the spectrum, corresponding to positive temperatures. 

Fig.~\ref{fig:fig3}a  confirms that the density of states is Gaussian and that it does not change much
as a moderate particle-particle interaction is introduced. 
Fig.~\ref{fig:fig3}b and c further show that in the noninteracting case, 
the probe temperature is very close to $T_{\rm th}$ of Eq.~\eqref{eq:temp}
except in the tail of the many-body density of states. This is due to well-known deviations from Gaussianity 
there~\cite{Bro81}.
For the case $m=16$ and $n=8$ shown in Fig.~\ref{fig:fig3}, 
the probe temperature differs from $T_{\rm th}$ by less than 10 \% for
$T_{\rm th}/\Delta \gtrsim 5.5$ for $U/\Delta=0$, corresponding to about 90 \% of the many-body eigenstates. 
A moderate interaction has very little impact, except in the middle of the spectrum, where large energy eigenvalues
fluctuations increase the discrepancy between the probe temperature and $T_{\rm th}$. 
We found that for $U/\Delta=0.2$, the difference between the two remains below 10 \% for 
$50 \gtrsim T_{\rm th}/\Delta \gtrsim 5.5$, corresponding to about 80 \% of the spectrum. These results
do not significantly change upon further increase of the interaction strength as long as $U/\Delta \ll 1$.

\section{From probe to eigenstate temperature}

The results presented so far seem to indicate that the probe temperature gives more than just an
operational temperature definition for individual many-body eigenstates. We now argue that this  
definition satisfies further properties expected of an absolute temperature~\cite{Huang}. First, it is clear from 
the currents definition, Eqs.~\eqref{eq:currents}, that the temperature definition does not depend on the magnitude 
of  the probe-system coupling amplitude $t$. This is so only as long as $t$ is small enough, so that
Eqs.~\eqref{eq:currents} are valid, but still leaves a sizeable range of parameter $t$. 

Second, the probe temperature is defined by the vanishing of the particle and heat
currents between the probe and the system, $I_A(\mu_A,T_A)=J_A(\mu_A,T_A) \equiv 0$. 
Let us 
introduce a second many-body state $|B\rangle$ and couple it to the probe while keeping the latter's
temperature and chemical potential the same. Let us further suppose then that there is no current between $|B\rangle$
and the probe, $I_B(\mu_A,T_A)=J_B(\mu_A,T_A) \equiv 0$. From 
Eqs.~\eqref{eq:currents}, one straightforwardly concludes that there would be no current either 
between $|A\rangle$ and $|B\rangle$, 
$I_{AB} = t^2 \sum_\alpha [f_A(\epsilon_\alpha) - f_B(\epsilon_\alpha)] = 0,$
$J_{AB} = t^2 \sum_\alpha  (\epsilon_\alpha-\mu) [f_A(\epsilon_\alpha) - f_B(\epsilon_\alpha))] = 0.$
The transitivity condition of the zeroth law of thermodynamics is thus satisfied by the probe
definition of the temperature. 

Third, in the regime of validity of Eqs.~\eqref{eq:currents}, the system may temporarily work as a heat engine
when the probe is biased away from the equilibrium condition, i.e. $T_A \rightarrow T_A + \delta T$ and
$\mu_A \rightarrow \mu_A + \delta \mu$. Assume that all fermions carry an electric charge $q$. With this bias,
both a heat and an electric current flow, with the latter being accompanied by electrical work. The efficiency
of the resulting heat engine is given by $\eta = - I_A \delta \mu/J_A$~\cite{Ben17} with, from  Eqs.~\eqref{eq:currents}
\begin{subequations}\label{eq:currents_bias}
\begin{eqnarray}
I_A (\delta \mu,\delta T)&=& -t^2 \sum_\alpha [\partial_\mu f_{\rm FD} \delta \mu
+\partial_T f_{\rm FD} \delta T] \, , \\
J_A (\delta \mu,\delta T)&=& - t^2 \sum_\alpha  (\epsilon_\alpha-\mu) [\partial_\mu f_{\rm FD} \delta \mu
+\partial_T f_{\rm FD} \delta T]\,  , \qquad
\end{eqnarray}
\end{subequations}
where in both expressions, $f_{\rm FD}=f_{\rm FD}(E_i=\epsilon_\alpha,\mu,T)$ and 
both $\delta T$ and $\delta \mu$ are assumed very small to justify the linearization 
of the Fermi-Dirac distributions. 
A straightforward calculation 
gives that the maximal efficiency of the engine is given by
\begin{eqnarray}\label{eq:zt}
\eta_{\rm max} &=& \left(\frac{\sqrt{1+\mathcal{ZT}}-1}{\sqrt{1+\mathcal{ZT}}+1}\right) \frac{
|\delta T|
}{T_A} \, .
\end{eqnarray}
The dimensionless figure of merit is given by
$\mathcal{ZT}^{-1}=\mathcal{L}^{(0)} \mathcal{L}^{(2)}/\left(\mathcal{L}^{(1)}\right)^2-1$, with
$ \mathcal{L}^{(0)}=\sum_\alpha \partial_\mu f_{\rm FD}$, 
$ \mathcal{L}^{(1)}=\sum_\alpha \partial_T f_{\rm FD} \, T_A$ and 
$ \mathcal{L}^{(2)}=\sum_\alpha (\epsilon_\alpha - \mu_A) \partial_T f_{\rm FD} \, T_A$.
Eq.~\eqref{eq:zt} defines a relative temperature scale in that 
$\eta_{\rm max}$  is a function of the temperature difference $\delta T$
between system and probe, in agreement with Carnot's theorem. \\

\section{Many-body eigenstate temperature}

There seem to be good reasons to take the probe temperature as a definition of the many-body eigenstate 
temperature. Nevertherless, a further condition needs to be satisfied before this is done. 
In general,  $I_B(\mu_A,T_A)=J_B( mu_A,T_A) \equiv 0$
still allows for partial currents $[f_A(\epsilon_\alpha) - f_{\rm FD}(E_i=\epsilon_\alpha,\mu,T)] \ne 0$
and $(\epsilon_\alpha-\mu) [f_A(\epsilon_\alpha) - f_{\rm FD}(E_i=\epsilon_\alpha,\mu,T)] \ne 0$
in Eqs.~\eqref{eq:currents}.
Therefore, 
the state of the few-fermion system will eventually change, even with a weak, finite system-probe coupling,
unless detailed balance conditions are  satisfied, 
\begin{eqnarray}\label{eq:dbal}
 [f_A(\epsilon_\alpha) - f_{\rm FD}(E_i=\epsilon_\alpha,\mu,T)] &=& 0 \, , \;\; \forall \alpha \, .
\end{eqnarray}
This of course means that particle occupancies in few-fermion states are given by the Fermi-Dirac distribution. 
Looking at Fig.~\ref{fig:fig1} we see that this may occur for sufficiently interacting few-fermion systems. 
To quantify the rate at which Fermi-Dirac like occupancies emerge as the interaction strength increases, 
we investigate the variance of the partial currents making up the particle
and heat currents of Eq.~\eqref{eq:currents},
\onecolumngrid
\begin{eqnarray}\label{eq:varcurrents}
\delta I^2_A = t^4 \sum_\alpha [f_A(\epsilon_\alpha) - f_{\rm FD}(E_i=\epsilon_\alpha,\mu_A,T_A)]^2 \; , \;\;\;\;
\delta J^2_A = t^4 \sum_\alpha  (\epsilon_\alpha-\mu_A)^2 [f_A(\epsilon_\alpha) - f_{\rm FD}(E_i=\epsilon_\alpha,\mu_A,T_A)]^2\, . \; 
\end{eqnarray}
\twocolumngrid
\noindent These current variances vanish only when the detailed balance conditions of Eq.~\eqref{eq:dbal}
are satisfied.
When this is the case, orbital occupancies in the few fermion system are given by the Fermi-Dirac
distribution, partial currents accordingly vanish and the many-body eigenstate $|A\rangle$ is at
equilibrium with the thermometric Fermi-Dirac gas in the usual sense. In particular, the weak coupling between 
the two subsystems does not change the state of the few fermion system.\\

\begin{figure}[h!]
\includegraphics[width=0.85 \columnwidth]{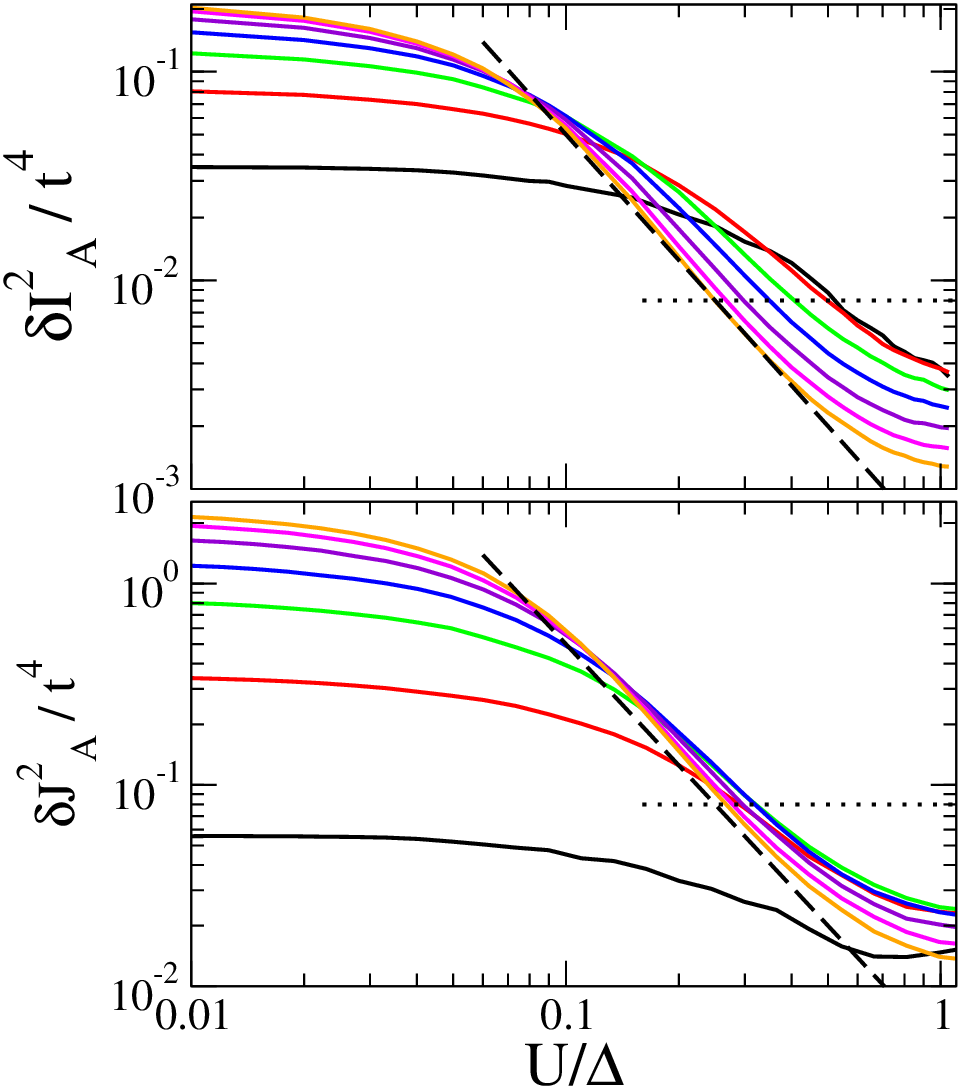}
\caption{\label{fig:fig4} (Color online) 
Deviation from detailed balance for the particle current (top panel) and the heat current (bottom panel) as a function of the 
interaction strength $U/\Delta$ for 1000 realizations of the two-body randomly interacting fermion model \eqref{eq:tbrim} with 
$m=12$ orbitals and $n=6$ fermions and thus $N=924$ many-body states. Excitations energies above the ground state
energy are $\delta E/\Delta=2.$ (black curve), 4. (red), 6. (green), 8. (blue), 10. (violet), 12. (magenta), and 14 (orange). 
Dashed lines indicate a power-law decay $\propto (U/\Delta)^{-2}$. 
The dotted line indicates the arbitrarily chosen threshold 
$\delta I_{A}^2/t^4 = 8 \times 10^{-3}$ and $\delta J_{A}^2/t^4 = 8 \times 10^{-2}$ used
to define critical interaction strengths $U_{c1}$ and $U_{c2}$.}
\end{figure}

Fig.~\ref{fig:fig4} shows $\delta I^2_A$ and $\delta J^2_A$ as a function of the 
normalized interaction strength $U/\Delta$ in the few fermion system. It is seen that as $U/\Delta$ increases,
both variances decrease with rates that depend on the excitation energy above the few fermion ground-state.
At high enough excitation energy -- though not necessarily too high -- we find that 
$\delta I^2_A, \delta J^2_A  \sim (U/\Delta)^{-2}$ until they saturate at a value depending on both 
the number of single-particle orbitals and of particles in the few fermion system. We have seen this 
behavior for other values of $m$ and $n$, not shown in Fig.~\ref{fig:fig4}. 

\begin{figure}[b!]
\centering
\hspace{-3mm}
\includegraphics[width=0.9 \columnwidth]{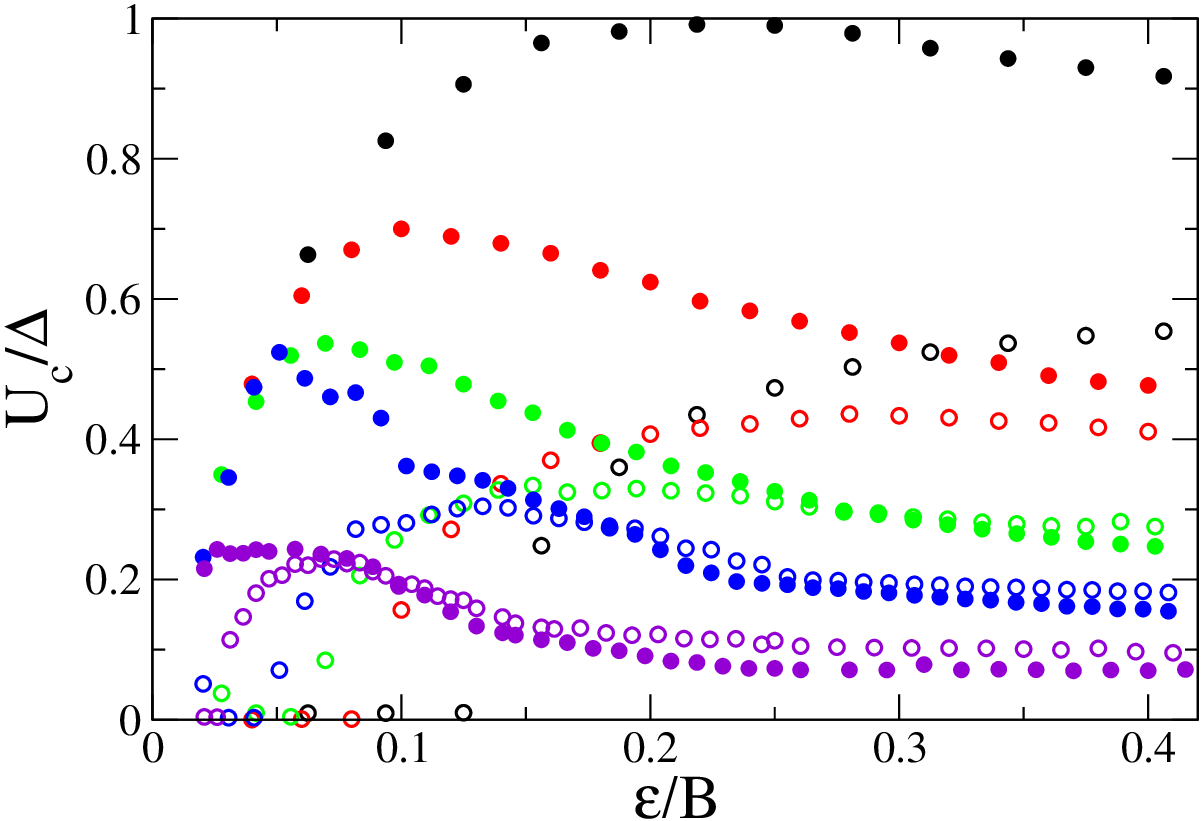}
\caption{\label{fig:fig5} (Color online) Critical interaction strength $U_c$ vs. normalized excitation energy $\epsilon/B$ with the many-body bandwidth $B\simeq n(m-n) \Delta$, 
for $m=8$, $n=4$ (black), $m=10$, $n=5$ (red), $m=12$, $n=6$ (green), $m=14$, $n=7$ (blue) and $m=16$, $n=8$ (violet). Solid circles give $U_{c1}$ and empty circles give $U_{c2}$ (see text).}
\end{figure}

We are interested in the parametric dependence of the 
rate at which the current variances vanish. To that end, we define critical interaction strengths $U_{c1}$ 
and $U_{c2}$ with
$\delta I_{A}^2(U_{c1}) = 8 \times 10^{-3} \, t^4$ and $\delta J_{A}^2(U_{c2})  = 8 \times 10^{-2} \, t^4$. We chose these
values, somehow arbitrary, because they correspond to interaction strengths with significantly reduced
current variances, but well before their large $U$, finite-size saturation for all cases considered, $m/n=2$
and $n = 4,5, ... 8$. 
Fig.~\ref{fig:fig5} shows the obtained values of $U_{c1}$ (solid circles) and $U_{c2}$ (empty circles)
as a function of the normalized excitation energy $\epsilon/B$. 
At small excitation energy close to the ground-state energy,  
both $U_{c1}$ and $U_{c2}$ increase with $\epsilon/B$. This reflects the fact that low excitations
effectively restrict the  number of available single-particle orbitals, which allows a faster transition to a
Fermi-Dirac distribution. Furthermore, quasiparticles with small excitation energies carry very little heat current, 
which explains why $U_{c2}$ is always very small at 
small excitation energies. For higher excitation energy,
both $U_{c1}$ and $U_{c2}$ are monotonously decreasing functions of the number of particles and orbitals, 
furthermore this holds above an excitation energy that also decreases with the number of particles. 

Because of the restricted range of variation of $n$ reachable by exact diagonalization, it is hard to extract  
a parametric dependence of $U_c$. Nevertheless, 
the data shown 
in Fig.~\ref{fig:fig5} at half filling, $n=m/2$ seem to indicate a behavior $U_{c1} \propto n^{-3}$, consistent
with the emergence of quantum chaos reported 
in Ref.~\cite{Jac97}. They also suggest that for sufficiently large systems, 
both particle and heat current based critical interactions become the same. 
With these data we conjecture that 
eigenstate thermalization is accompanied by ensemble equivalence, where each individual few-fermion 
eigenstates exhibits a Fermi-Dirac occupancy distribution, and accordingly defines a canonical ensemble,
above a critical interaction strength $U_{c} \propto n^{-3}$ for systems at half filling. 
For $n=8$, this already corresponds to a rather
weak interaction strength, $U_c \lesssim 0.1 \Delta$. 

\section{Conclusion}

We have shown that few fermion systems have eigenstates with a Fermi-Dirac occupancy of single-particle
orbitals, provided they have a sufficiently strong interaction. Our results indicate that, not too close to the ground-state
energy, the critical interaction
strength scales parametrically as $U_c \propto n^{-3}$ in systems with constant filling factor. In particular, 
$U_c \lesssim 0.1 \Delta$ with the single-particle orbital  spacing $\Delta$, for $n=8$ fermions on $m=16$ orbitals. 
This indicates that for still small systems with, say, $n=20$, the critical interaction strength  
is only a fraction of this single-particle orbital spacing. Ensemble equivalence at the level of individual 
few-body eigenstates is therefore achieved very fast in the number $n$ of fermions, and requires only a
weak interaction. 

This work has been supported by the Swiss National Science Foundation.

\end{document}